\documentclass[prb,a4paper,twocolumn,showpacs,superscriptaddress]{revtex4}
\usepackage{amsmath}
\usepackage{amsfonts}
\usepackage{amssymb}
\usepackage{bm}
\usepackage{graphicx}
\usepackage[colorlinks]{hyperref}

\begin{document}

\title{Effect of decoherence on resonant Cooper-pair tunneling in a voltage-biased single-Cooper-pair transistor}
\author{J. Lepp\"akangas}
\email[Electronic address: ]{juha.leppakangas[at]oulu.fi}
\affiliation{Department of Physical Sciences,
P.O.Box 3000, FI-90014 University of Oulu, Finland}
\author{E. Thuneberg}
\affiliation{Department of Physical Sciences,
P.O.Box 3000, FI-90014 University of Oulu, Finland}

\date{\today}


\begin{abstract}

We analyze how decoherence appears in the
$I-V$ characteristics of a voltage-biased single-Cooper-pair transistor. 
Especially the effect on resonant single or several
Cooper-pair tunneling is studied.
We consider both a symmetric and an asymmetric transistor.
As a decoherence source we use a small resistive impedance
($\textrm{Re}[Z(\omega)]\ll R_Q=h/4e^2$) in series with the transistor,
which provides both thermal and quantum fluctuations of the voltage. Additional
decoherence sources are quasiparticle
tunneling across the Josephson junctions and quantum $f$-noise caused by
spurious charge fluctuators nearby the island.
The analysis is based on a real-time diagrammatic technique which includes Zeno-like
effects in the charge transport, where the tunneling is slowed down due to
strong decoherence.
As compared to the
Pauli-master-equation treatment of the problem, the present results are more consistent with experiments where many of the predicted sharp resonant structures are missing or weakened due to decoherence.
\end{abstract}

\pacs{03.65.Yz, 73.23.Hk, 74.50.+r, 74.78.Na, 85.25.Cp}

\maketitle

\section{Introduction}\label{introduction}

The quantum mechanical effects originating in coherent tunneling of Cooper pairs
in small Josephson junctions (JJ) have been investigated actively in recent
years due to their
great potential to be used in nanotechnological applications in the future~\cite{qengineer,qbits}. A central obstacle has
been that the effects are easily decohered
by an uncontrolled coupling between the studied system and its nearby environment~\cite{qubitnoise}.
However, due to their high sensitivity to the environmental
fluctuations, the small JJs can as well be used as
probes of physics in mesoscopic low temperature devices~\cite{ingold3,ingold,sonin,pekola}.

In this paper we study theoretically how decoherence appears in the
$I-V$ characteristics of a voltage-biased single-Cooper-pair transistor~\cite{qengineer,fulton} (SCPT).
The noise in the source-drain line of the SCPT is dominated by quantum mechanical
fluctuations of voltage due to finite transmission line impedance, or more generally due to coupling to the electromagnetic environment~\cite{joyez} (EE).
The noise in the gate (charge) is due to spurious charge fluctuators~\cite{nakamura1,faoro1} (CF) in the materials surrounding the island, as the EE noise in this line is usually shielded by a small gate capacitance.
These decoherence sources are accompanied by quasiparticle tunneling across the JJs, where a single electron tunnels across a JJ
with the cost of breaking a Cooper pair. In this work we model the SCPT's
transport properties to the second order
in the interaction with the relevant unperturbed baths describing EE, CF and quasiparticles.
This leads to the demand that the resistance associated with the EE and CF has
to be small ($\ll R_Q=h/4e^2$) and that the tunneling resistances of the JJs have to be large ($>R_Q$).

We study the $I-V$ characteristics in the subgap regime where resonant tunneling
of Cooper pairs is the essential conduction mechanism~\cite{brink1,brink2,haviland}.
The average current in this regime has been previously calculated by
using the Pauli master equation~\cite{brink1,brink2,joyez,siewert1} modeling
the populations of the SCPT's eigenstates.
We call this treatment as the coherent tunneling model (CTM).
Practically every tunneling event of one or more Cooper pairs across the JJs in the SCPT can be made resonant by properly adjusting the
transport and gate voltages. It follows that the CTM-current
has quite cumbersome features as the subgap region is full of peaks
with alternating heights and widths. Experimentally, however, the first-order (single particle) resonances are well visible~\cite{haviland} but only few of the higher-order (several particle) resonances have been identified~\cite{haviland,joyez,fitzgerald,billangeon}.
This kind of ``wash-out'' of the higher-order resonances
occurs apparently due to strong decoherence caused by the environment as
compared to the widths of the resonances. The results obtained by the CTM
are valid only for the strongest resonances and therefore only qualitative
match between this theory and experiments can be obtained.

We model the system
using a density matrix approach (DMA) based on a Keldysh type real-time diagrammatic technique~\cite{qengineer,keldysh1} in the Born approximation (leading order). As compared to the CTM, the method takes into account
also nondiagonal contributions in the master
equation. This enables Zeno-like effects~\cite{zeno2,zeno} in the charge transport, where strong decoherence
drives, or continuously ``measures'', the SCPT into superpositions of eigenstates that are more stable
under the decoherence. In Ref.~[\onlinecite{zeno}] this was used to analyze the experimental findings of Ref.~[\onlinecite{bibow}] for a double-island
system using the EE as the relevant source of decoherence.
Also the charge tunneling nearby the so-called JQP-cycles has widely been analyzed by similar methods~\cite{choi1,choi2,clerk,johansson1,johansson2} considering quasiparticle tunneling as a perturbation.
This paper extends these results in the sense that the method applies to arbitrary resonances
and includes all the relevant noise sources in the voltage-biased SCPT. We show by numerical simulations that in typical experimental
conditions most of the higher-order resonances are indeed lost, unless a careful
filtering of the relevant noise is done.
We also show that the ohmic environments (EE and CF) leave different
traces to the $I-V$ characteristics
so that one can, in principle,
identify the strength of EE and CF separately in a specific experimental setup.
In the case of asymmetric SCPT and strong dissipation,
the model produces the results of Ref.~[\onlinecite{oma}]
(incoherent tunneling of Cooper pairs across the small JJ). In the opposite limit of weak dissipation, the model produces the results of Ref.~[\onlinecite{lt24}] obtained by the CTM.

The paper is organized as follows.
In section \ref{model} we introduce the model Hamiltonian for the SCPT
and discuss its properties. In section \ref{keldysh} we go through the diagrammatic technique used in the modeling, the environmental contribution and point out the problems that might occur
in the DMA and how to circumvent them.
The Appendix is devoted to specifics of this treatment.
In section \ref{tracing} we show how average properties can be
calculated by Laplace-transforming and tracing the equation of motion.
Section \ref{results} shows and analyzes numerical solutions of the problem.
The conclusions are given in section \ref{conclusion}.

\section{The SCPT Hamiltonian}\label{model}

\begin{figure}[tb]
\begin{center}\leavevmode
\includegraphics[width=\linewidth]{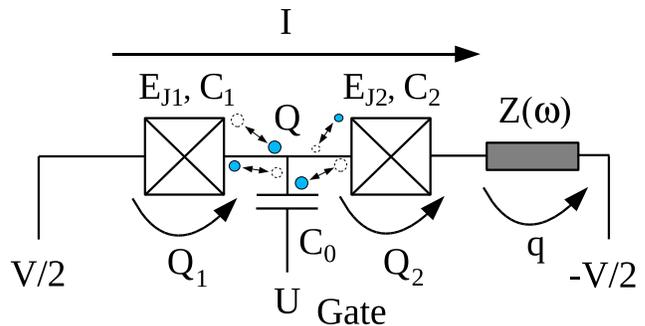}
\caption{(Color online) A voltage-biased SCPT consisting of two JJs (crossed boxes)
in series with the voltage source $V$. The island is capacitively coupled
to a gate lead. Also is drawn the EE described by an impedance $Z(\omega)$
and spurious charge fluctuators nearby the island (section \ref{keldysh}).}
\label{kuvasset}
\end{center}
\end{figure}

We model Cooper-pair tunneling across the JJs in a nonperturbative way so that the calculation automatically includes Cooper-pair tunneling processes
in arbitrary order.
A schematic drawing of the SCPT is shown in Fig.~\ref{kuvasset}.
The starting point is the Hamiltonian~\cite{brink1}
\begin{gather}
H_{\rm SCPT}=\frac{\left(Q_1-Q_2+Q_0\right)^2}{2C_{\Sigma}}-\frac{V}{2}(Q_1+Q_2)\nonumber\\
-E_{J1}\cos(\varphi_1)-E_{J2}\cos(\varphi_2),
\label{ham}
\end{gather}
where $Q_i$ is the charge tunneled across the $i$:th JJ,
a conjugated variable to the phase difference $\varphi_i$, and $E_{Ji}$ is the
related Josephson coupling energy. The capacitance of the $i$:th JJ is $C_i$, the
capacitance of the gate $C_0$, $C_{\Sigma}=C_0+C_1+C_2$
and $Q_0=C_0U+(C_1-C_2)V/2$ is the quasicharge.
In the absence of quasiparticle tunneling the operator $Q_{i}$ has $2e$ quantization
or, equivalently, the eigenfunctions in the $\varphi$-space are $2\pi$-periodic~\cite{period}.
Depending whether we are considering an almost symmetric or a highly asymmetric SCPT, there
are now two convenient bases for calculations.

\subsection{Symmetric SCPT and no quasiparticle tunneling}\label{model1}

For a symmetric SCPT (identical JJs) the Hamiltonian is most easily solved in the
island and feed charge basis, which is defined as
$Q=Q_1-Q_2$ and $\bar{Q}=Q_1+Q_2$. The Hamiltonian can now be written as
\begin{gather}
H_{\rm SCPT}=\frac{\left(Q+Q_0\right)^2}{2C_{\Sigma}}-\frac{1}{2}\bar QV\nonumber\\
-\frac{E_{J}}{2}\left(  T_QT_{\bar{Q}}+T^{\dagger}_QT_{\bar{Q}}+T_QT^{\dagger}_{\bar{Q}}+T^{\dagger}_QT^{\dagger}_{\bar{Q}}  \right),
\label{hamsym}
\end{gather}
where we have defined the charge (Cooper pair) translation operators
$T_Q=\sum_{Q}\vert Q+2e\rangle\langle Q\vert$,
$T_{\bar Q}=\sum_{\bar Q}\vert \bar Q+2e\rangle\langle \bar Q\vert$ and
$E_J=E_{J1}=E_{J2}$.
One sees that if $\vert j\rangle$ is an eigenstate with an eigenenergy $E$, then the
translated state $T_{\bar Q}^{(\dagger)}\vert j\rangle$ is also an eigenstate with an eigenenergy $E-eV$ ($E+eV$).
Such states are called equivalent but belong to different steps of Wannier-Stark
ladder called zones.
The product state basis $\vert Q\rangle\vert\bar{Q}\rangle$ is the
most convenient one for a numerical solution in the case $E_J<E_C=e^2/2C_{\Sigma}$~\cite{brink1}.
Since elementary tunneling processes change $Q_i$ by $2e$, there is no coupling
between $\vert 0,0\rangle$ and $\vert 0,2e\rangle$,
where we now use the notation $\vert Q,\bar{Q}\rangle$ for the
product state $\vert Q\rangle\vert\bar{Q}\rangle$. Therefore one can halve the
dimension of the Hamiltonian matrix by leaving out the state $\vert 0,2e\rangle$,
and all the other states obtained from this by the elementary tunneling processes.
It follows that the zones differ from each another by a
translation of $4en$ of the feed charge and by $-2eVn$ in the eigenenergy,
where $n$ is an arbitrary integer.

\subsection{Asymmetric SCPT and no quasiparticle tunneling}\label{model2}

The other special limit is the asymmetric SCPT~\cite{rene,oma} where by
asymmetry we mean that $E_{J1}\gg E_{J2}$, which usually leads to $C_1\gg C_2$ also.
The circuit can be seen
as a Cooper-pair box~\cite{qengineer} (CPB) which is probed, or excited, by the smaller JJ (probe).
A convenient basis for this case is defined as
$Q=Q_1-Q_2$, $Q_{\Sigma}=Q_2$, $\varphi=\varphi_1$
$\varphi_{\Sigma}=\varphi_1+\varphi_2$. This canonical
transformation leads to the Hamiltonian
\begin{gather}
H_{\rm SCPT}=\frac{\left(Q+Q_0'\right)^2}{2C_{\Sigma}}-\frac{E_{J1}}{2}\left( T_Q+T^{\dagger}_Q  \right)\nonumber\\
-Q_{\Sigma}V-\frac{E_{J2}}{2}\left( T_{Q}T^{\dagger}_{Q_{\Sigma}}+T^{\dagger}_{Q}T_{Q_{\Sigma}}  \right),
\label{hamasym}
\end{gather}
where $Q_0'=C_0U-C_2V$ (we have assumed that $C_0\ll C_{2}$). The first two terms
on the r.\ h.\ s.\ of Eq.~(\ref{hamasym}) correspond to the CPB Hamiltonian and the last two terms describe single tunneling
events of Cooper pairs across the probe with simultaneous excitation of the CPB. The eigenstates of the CPB are mixed by the probe tunneling part
mostly when $2eVn$, where $n$ is an arbitrary integer,
matches certain energy level difference of the CPB.

We solve the eigenstates of the asymmetric SCPT representing the
Hamiltonian in the product state basis
$\vert j,Q_0'\rangle\vert Q_{\Sigma}\rangle$,
where $\vert j,Q_0'\rangle$ is the $j$:th eigenstate of the CPB Hamiltonian,
which has to be calculated separately, corresponding to the quasicharge $Q_0'$.
The resulting SCPT eigenstates have the Wannier-Stark structure with steps separated by a translation $2en$ of the charge $Q_{\Sigma}$
and a change $-2eVn$ in the eigenenergy.

\subsection{Quasiparticle tunneling}

The preceding treatment is valid when no quasiparticle tunneling
exists, or its effect can be neglected. In the opposite case the quantization of the tunneled charge $Q_i$
must be changed from $2e$ to $e$.
We treat the quasiparticle tunneling perturbatively causing transitions between unperturbed states (section \ref{qparticles}).
Since the unperturbed Hamiltonian $H_{\rm{SCPT}}$ describes only Cooper-pair-tunneling processes, its eigenstates are superpositions of
states differing $2ne$ in the tunneled charges, where $n$ is an arbitrary
integer. In other words the Hamiltonian matrix is a sum of four
blocks each operating in a subspace possessing
a different parity combination (even or odd) of the tunneled charges $Q_i/e$.
The eigenfunctions in each of the subspaces can be solved independently
by using the same Hamiltonian as before, but just properly shifting the
island and/or the feed charge.
Only subspaces with different parity of the island charge have different
structures.

\section{Effect of the environment}\label{keldysh}

The interaction between the SCPT and its environment
leads to dissipative (open) quantum mechanics~\cite{weiss} and to a net current across the system. Each
dissipation mechanism shows up in a characteristic way.
The noise in the voltage across the SCPT is due to a finite transmission line impedance (EE) described by $\textrm{Re}[Z(\omega)]$, which we assume to be a
constant $R$ at low frequencies.
The noise in the gate charge is due to the CF, which
also can be fully characterized by an analogous resistance $R_{\rm{CF}}$
if the quantum mechanical noise spectrum is of $f$-type~\cite{nakamura1,faoro1}
(and we deal with zero temperature).
These decoherence sources are accompanied by quasiparticle tunneling across the JJs.
We include
the interaction with the relevant baths by using a real-time Keldysh diagrammatic technique~\cite{keldysh1,keldysh2,keldysh3}. This leads to a
master equation for the SCPT's density matrix as the environment is
traced out of the equations by using the assumption of unperturbed
reservoirs. We restrict the analysis to the second-order calculation in the
interaction with each of the baths since the higher-order calculation turns out to be actually less suitable for a reliable analysis (Appendix).
This is justified for the EE (CF)
for which $R_{({\rm{CF}})}\ll R_Q$ and for low transparency JJs with tunneling resistances $R_T>R_Q$.

\subsection{Coupling to the electromagnetic environment}\label{EE}

For the case of a dissipative EE,
the phenomenological total Hamiltonian describing a voltage-biased SCPT in series with the impedance can be presented in the form~\cite{zeno,ingold3}
\begin{gather}
H_{\rm{total}}=H_{\rm{SCPT}}-Q_{\rm{int}}\frac{q}{C_{\rm{int}}}+\frac{Q_{\rm{int}}^2}{2C_{\rm{int}}}\nonumber\\
+\frac{q^2}{2C_{\rm{int}}}+\sum_{l}\left[\frac{q_l^2}{2C_l}+\frac{\hbar^2}{e^2}\frac{(\varphi_l-\varphi_R)^2}{2L_l}\right],
\label{dis}
\end{gather}
where the fluctuations caused by the EE are modeled by coupling an infinite
number of $LC$-oscillators to the phase difference
variable $\varphi_R$ across the impedance.
The EE couples to the SCPT
through the term $-Q_{\rm{int}}q/C_{\rm{int}}$, where the charge $q$ passed through the voltage source (relative to the equilibrium charge $C_{\rm{int}}V$)
is a conjugate variable to $\varphi_R$.
The $l$-sum part in Eq.\ (\ref{dis}) can be shown to describe correctly the
fluctuations across any dissipative impedance
by properly choosing the parameters $L_l$ and $C_l$. The extra charging part, which can
be written in the form $(Q_{\rm{int}}-q)^2/2C_{\rm{int}}$, can be understood
as the capacitive energy of the SCPT seen by the impedance.
The Kirchhoff rules then lead to the identifications (in the case of the EE)
\begin{gather}
Q_{\rm{int}}=\frac{C_2Q_1+C_1Q_2}{C_{\Sigma}}=\frac{C_2}{C_{\Sigma}}Q+Q_{\Sigma}\equiv Q_{\rm{int}}^{\rm{EE}},
\label{interaction}
\end{gather}
and $C_{\rm{int}}=(1/C_1+1/C_2)^{-1}\equiv C_{\rm{int}}^{\rm{EE}}$ (assuming that $C_0\ll C_1,C_2$). This representation of the fluctuations corresponds to series $LC$-oscillators in
parallel with the SCPT. Alternatively, the EE could also be modeled as parallel $LC$-oscillators in series~\cite{joyez}.

In the real-time diagrammatic technique the EE is assumed to be a large reservoir which is not significantly affected by the
tunneling in the SCPT and is in a good approximation described by the last two terms in Eq.\ (\ref{dis}).
In this case the environmental part can be
traced out and described only via its equilibrium properties given by the quantum fluctuation-dissipation theorem
\begin{gather}
\alpha(t)=\alpha(-t)^{\ast}=\langle\delta V(t)\delta V(0)\rangle\nonumber\\
=\frac{R}{\pi}\int_{-\infty}^{\infty}\frac{\hbar\omega}{1+\left(\frac{\omega}{\omega_c}\right)^2}\frac{e^{-i\omega t}}{1-e^{-\beta\hbar\omega}}d\omega,
\label{alpha}
\end{gather}
where $\delta V=q/C_{\rm{int}}$, $\omega_c=1/(RC_{\rm{int}})$ and $\beta=1/k_bT$. The calculation
now reduces to rules for creating diagrams and the corresponding
generalized transition rates (Appendix). The first order
diagrams lead to the master equation~\cite{zeno}
\begin{gather}
\dot\rho(t)=L_0\rho(t)+\int_{t_0}^{t}\Sigma(t-t')\rho(t')dt'+L_N\rho(t),
\label{density1}
\end{gather}
where $L_0=(i/\hbar)[\cdot,H_{\rm{SCPT}}]$ is the Liouville operator,
$\Sigma(t)$ the generalized transition rate and $L_N=(i/\hbar)[\cdot,Q_{\rm{int}}^2/2C_{\rm{int}}]$ its renormalization. The transition rate tensor is given by
\begin{gather}
\Sigma(t)=\alpha'(t)L_{\rm{int}}e^{L_0t}L_{\rm{int}}-i\alpha''(t)L_{\rm{int}}e^{L_0t}M_{\rm{int}},
\label{1order}
\end{gather}
where $\alpha(t)=\alpha'(t)+i\alpha''(t)$,
$L_{\rm{int}}=(i/\hbar)[\cdot ,Q_{\rm{int}}]$ and
$M_{\rm{int}}=(i/\hbar)[\cdot ,Q_{\rm{int}}]_+$. The renormalization $L_N$
can be shown  to cancel certain terms in Eq.\ (\ref{1order}) (Appendix) and to make the
renormalized transition rate $\tilde\Sigma(t-t')=\Sigma(t-t')+\delta(t-t')L_N$ invariant
under a $4e$ translation of
$\bar Q$, or $2e$ translation of $Q_{\Sigma}$ in the asymmetric case.

\subsection{Coupling to the charge fluctuators}\label{CF}

The CF can be described similarly as the EE since we assume its noise to be of $f$-type.
This means that its noise spectral density is proportional to the frequency
and that $T_{\rm{CF}}=0$ (temperature of the CF).
Also certain effective nonvanishing value of $T_{\rm{CF}}$ could be used for describing dephasing due to thermal fluctuations of the CF. However, in this case the model would not be ``universal'' since an ensemble of harmonic oscillators produces different
kind of thermal noise as, for example, an ensemble of two-level systems.
In the description
the island charge $Q=Q_1-Q_2$ is identified as $Q_{\rm{int}}$ of the CF. It is coupled linearly to the fluctuating
operator $q/C_{\rm{int}}^{\rm{CF}}$, which has the desired
properties i.~e.\ ohmic dissipation characterized by $R_{\rm{CF}}$ and a cutoff frequency
$\omega_c=1/R_{\rm{CF}}C_{\rm{int}}^{\rm{CF}}$. Also the renormalization term $Q^2/2C_{\rm{int}}^{\rm{CF}}$ has to be introduced in order to compensate the change in the effective charging potential felt by the SCPT due to the embedded CF~\cite{weiss,caldeira}. As a result one ends up with the same total Hamiltonian as in Eq.\ (\ref{dis}).
Therefore the master equation (\ref{density1}) applies also to the CF but now with the identifications
$Q_{\rm{int}}=Q_{\rm{int}}^{\rm{CF}}\equiv Q$, $T=T_{\rm{CF}}$, $R=R_{\rm{CF}}$
and $C_{\rm{int}}=C_{\rm{int}}^{\rm{CF}}$.
In the simulations we use the value $C_{\rm{int}}^{\rm{CF}}=C_{\rm{int}}^{\rm{EE}}$.

\subsection{Quasiparticle tunneling}\label{qparticles}

The quasiparticle tunneling across the JJs is described
by using the tunneling-Hamiltonian formalism~\cite{tunnel1,tunnel2}.
The interaction operator between the quasiparticle reservoirs and the SCPT is of the form
\begin{gather}
H_{Ti}=T_{Q_i,e}  \otimes\sum_{p\in L,k\in R }t_{p,k}^ic_p^{\dagger}c_k+\rm{h.c.},
\label{qperturbation}
\end{gather}
where $c_j^{(\dagger)}$ is the electron annihilation (creation) operator of
the state $j$ in the left (L) or right (R) hand side lead (quasiparticle reservoir) of the $i$:th JJ, $t_{l,r}^i$ the related tunneling amplitude and
$T_{Q_i,e}=\sum_{Q_i}\vert Q_i+e\rangle\langle Q_i\vert$.
The tunneling rates are calculated
using the second-order diagrammatic treatment
and dropping out the terms that lead to $2e$-tunneling of charge (Josephson effect).
The environmental contribution is now traced in a correlation function
\begin{gather}
\alpha(t)_{\rm{qp}}^i=\sum_{k,l}\vert t_{k,l}^i\vert^2f(E_l)[1-f(E_k)]e^{it(\omega_l-\omega_k)},
\label{qp}
\end{gather}
where the sum over the states $k$ and $l$ is done in the equivalent semiconductor picture~\cite{tinkham} and $f(E)$ is the Fermi-distribution function.
No cut-off has to be introduced since we include only the
real part of the Laplace transformed correlation function (Appendix),
as the imaginary part would only lead
to a small correction on the island's capacitance in the case of
low transparency JJs~\cite{aleshkin,keldysh2}.
The factor $\alpha(t)_{\rm{qp}}^i$ can be related to the well known superconductor-superconductor quasiparticle current~\cite{tinkham}
and the operators $T_{Q_i,e}$ give the dependence on the tunneling
direction via the corresponding diagram rules (section \ref{lindblad}).

According to the model above, the equilibrium $I-V$ characteristics are always $e$-periodic
in $Q_0$, even with zero subgap conductance, and it seems that quasiparticles
can never be neglected. However, the average time between the switchings of different parity
states of the island charge via higher order processes can become very long.
In Refs.~[\onlinecite{odddecay,siewert1}] the decay of the odd parity state
(unpaired electron tunnels off the island) was introduced to restore the $2e$-periodicity in this situation.
We include the effect by introducing (on the top of the ordinary quasiparticle tunneling rate) a constant rate
\begin{gather}
\gamma_{\rm{esc}}\approx\frac{1}{2e^2R_T^iN_I}\theta(E_a-E_m),
\end{gather}
for the tunneling of the unpaired electron across the $i$:th JJ~\cite{siewert1}.
Here $N_I$ is the island's density of states, $\theta(x)$ the step function, and $\vert a\rangle$ and $\vert m\rangle$ are
the initial (odd parity) and final states of the corresponding Lindblad-equation
(section \ref{lindblad}). Theoretically the rate should be enhanced when
$E_a\gtrsim E_m$ due to singular quasiparticle density of states, but
we neglect this effect. Throughout the paper we use the value $1/N_I=10^{-2}$ $\mu$eV.

\subsubsection{The Lindblad form of quasiparticle tunneling}\label{lindblad}

In the case of quasiparticle tunneling (\ref{qperturbation}) and in some parameter region, the numerical solution for the steady state of the reduced density matrix fails by leading to
negative probabilities for occupancies
[this can be studied by solving Eq.\ (\ref{main}) for
$k=0$ in the limit $s\rightarrow 0$].
In these regions the calculation
of the average current naturally fails also. This physical inconsistence is
possible since our truncated (and therefore more or less phenomenological~\cite{loss}) equation is not of Lindblad form.
To circumvent the problem a proper
Lindblad approximation of the master equation, which does not change the results in the ``safe'' regions too much, should be used in the case of quasiparticle tunneling.

The Lindblad form can be obtained by two reasonable
approximations. The first one is the so-called Markovian assumption,
which in general says that the transition rate at time $t$
does not depend on the history $\rho(t')$, $t'<t$. In this context
one assumes that
\begin{gather}
\Sigma(t-t')\rho(t')\approx\nonumber\\
\Sigma(t-t')\exp[iH_0(t-t')]\rho(t)\exp[-iH_0(t-t')],
\end{gather}
i.~e.~the density matrix is approximately a constant in the interaction picture
on the timescale where the correlation functions die out.
In the diagram language this leads to one extra rule (Appendix) which makes the
first order equations equivalent to the so-called Born-Markov (BM) equations~\cite{weiss}
in the limit $t_0\rightarrow -\infty$.
As the environment is not dependent on the evolution of SCPT,
the Markov approximation is valid since the memory decay time
is small compared to the timescale of transitions or oscillations
of the superpositions~\cite{clerk}.

In the second step one redefines the nondiagonal transition rates
in the Lindblad form.
The positive-direction quasiparticle transition rates across the $i$:th JJ 
are in the BM-approximation
\begin{gather}
\Sigma^{a\rightarrow m}_{b\rightarrow n}=T_{ma}^i(T_{nb}^i)^*\left[ \frac{\Gamma^i(E_{a}-E_{m})}{2}+\frac{\Gamma^i(E_{b}-E_{n})}{2} \right],
\label{rate1}
\end{gather}
where $T_{ma}^i$ is the matrix element of the operator $T_{Q_i,e}$ and $\Gamma^i(E)=I_{\rm{q-q}}(E/e)/e$ is the classical quasiparticle
transition rate across the $i$:th JJ. The negative direction tunneling is obtained by
the substitution $T_{Q_i,e}\rightarrow T_{Q_i,e}^{\dagger}$.
The related decay rates (the second-type diagrams) are
\begin{gather}
\Sigma^{a\rightarrow m}_{b\rightarrow n}=-\frac{1}{2}\sum_v[(T_{mv}^i)^*T_{va}^i\Gamma^i(E_{a}-E_{v})\delta_{bn}\nonumber\\
+T_{nv}^i(T_{vb}^i)^*\Gamma^i(E_{b}-E_{v})\delta_{am}].
\label{rate2}
\end{gather}

A suitable Lindblad form of the density matrix equation is then a sum of
four independent operations
\begin{gather}
\bar\Sigma\rho=\sum_{i,\pm}\left[U_{i,\pm}\rho {U^{\dagger}_{i,\pm}}-\frac{1}{2}({U^{\dagger}_{i,\pm}}U_{i,\pm}\rho+\rho {U^{\dagger}_{i,\pm}}U_{i,\pm})\right]
\label{rate3},
\end{gather}
where $U_{i,+}$ is chosen to be
\begin{gather}
U_{i,+}=\sum_{ma}T^i_{ma}\sqrt{\Gamma(E_{a}-E_{m})}\vert m\rangle\langle a\vert,
\end{gather}
and the operator $U_{i,-}$ is obtained by the substitution
$T_{Q_i,e}\rightarrow T_{Q_i,e}^{\dagger}$.
This gives the same diagonal elements as the sum of contributions~(\ref{rate1})
and (\ref{rate2}) but redefines, for example, the nondiagonal rates of Eq.~(\ref{rate1}) to be
\begin{gather}
\bar\Sigma^{a\rightarrow m}_{b\rightarrow n}=T_{ma}^i(T_{nb}^i)^*\sqrt{\Gamma^i(E_{a}-E_{m})}\sqrt{\Gamma^i(E_{b}-E_{n})}.
\label{rate4}
\end{gather}
Now, if using $\bar\Sigma$, the transition rates that include forbidden diagonal transitions
(i.~e.~for which $E_a-E_{a'}<2\Delta$) are lost, but other processes are present with
approximately the same rate due to
the high gap energy $\Delta$. In more detail, by assuming that $E_{a}-E_{m}>E_{b}-E_{n}>2\Delta$, the relevant rates satisfy $\Gamma(E_{a}-E_{m})/\Gamma(E_{b}-E_{n})\approx (E_{a}-E_{m})/(E_{b}-E_{n})\ll 10$, from which it follows that the Eqs.\ (\ref{rate1}) and (\ref{rate4}) are
almost equivalent. Numerical simulations show that in the parameter region where
the density matrix calculated from the original master equation does not lose its positivity,
the above Lindblad form leads approximately to the same $I-V$
characteristics. It then removes the negativity of the stationary state solutions in the problematic regions.

\subsubsection{Quasiparticle tunneling thresholds}\label{treshold}

There exists a discontinuous jump in the theoretical quasiparticle tunneling
rates as a function of the voltage since the subgap current occurs only via thermally excited quasiparticles.
The BCS subgap resistance is approximately a constant\cite{brink1}
$R_q=R_T\exp(\Delta/k_bT-1.76)$, which means that the current is neglible for typical parameter values.
However, in experiments the thresholds broaden to smooth steps and
a finite subgap resistance persists, probably due to material imperfections~\cite{balatsky,toppari}.
One extra reason for the threshold broadening in the SCPT circuit is the thermal fluctuations of the voltage (due to the EE).
To include this effect one can proceed similarly as in the
$P(E)$-theory~\cite{ingold}.

The probability function [$P(E)$-function]
describing incoherent quasiparticle tunneling in the case of a single JJ, small environmental resistance $R$,
and finite temperature $T$ is at low energies
(the inhomogeneity of the integral equation in Ref.~[\onlinecite{ingold2}])
\begin{gather}
P_{e}(E)\approx\frac{2}{\pi}\frac{\Gamma_{e}}{\Gamma_{e}^2+4E^2},
\end{gather}
where $\Gamma_{e}=4\pi k_bTR/R_K$ and $R_K=h/e^2$. The $P_e(E)$-function is damped by the Boltzmann factor $e^{-E/k_bT}$ for larger negative values of $E$. The EE absorbs ($E>0$) or emits ($E<0$) energy $E$ with this probability
in the tunneling process. For Cooper-pair tunneling one
simply has to change $R_K$ to $R_Q=R_K/4$,
which then leads to a width $\Gamma_{2e}=4\Gamma_{e}$
of the $P_{2e}(E)$-function. Generally the tunneling of the charge $ne$
leads to the $P_{ne}(E)$-function which is obtained by replacing $R_K$ by $R_K(1/n)^2$
and therefore has the width $\Gamma_{ne}=\Gamma_{e}n^2$.

To utilize this for a SCPT, we notice that the charge seen by the impedance is
$(C_2Q_1^{\rm{real}}+C_1Q_2^{\rm{real}})/C_{\Sigma}$ (where $Q_i^{\rm{real}}$ is the charge of the capacitor $i$). The change of this variable, using the relevant
initial and final states, determines the amount of charge
transferred across the impedance in
the decay process (in order to regain the voltage balance), which then gives the correct value for $n$.
For example, in the case of an ordinary quasiparticle tunneling in a symmetric SCPT, it changes by $e/2$.
For the asymmetric SCPT ($E_{J1}\gg E_{J2}$) but still with identical capacitances ($C_1=C_2$)
and $E_{J1}\gg E_C$, the incoherent Cooper-pair tunneling across the probe leads to a change $2e$ since the CPB also immediately balances
the voltage across the larger JJ to zero on the average (the energy of the final state has no
quasicharge dependence). In the limit $E_{J1}\ll E_C$ this does not occur and the charge changes by $e$.
The $P(E)$-theory gives then for the ``corrected'' quasiparticle tunneling rates [to be used in Eq.~(\ref{rate3})]
\begin{gather}
\tilde {\Gamma}^i(E_a-E_m)=\int_{-\infty}^{\infty}P_{ne}(E')\Gamma^i(E_a-E_m-E')dE',
\end{gather}
where $n$ depends on the states $\vert a\rangle$ and $\vert b\rangle$
and on the nature of the process. This changes the thresholds from sharp
to smooth steps.

\section{Calculation of the current}\label{tracing}

In studying the average properties, one can simplify the
density matrix analysis by Laplace-transforming
the equations ($\int_{t_0}^{\infty}dte^{-s(t-t_0)}$) and considering the limit $s\rightarrow 0$. The transformation changes the master equation (\ref{density1}) to an algebraic equation
\begin{gather}
s\rho(s)-\rho^{t_0}=\tilde\Sigma(s)\rho(s).
\label{laplace2}
\end{gather}
Here $\tilde\Sigma(t)$ is the (renormalized) transition rate tensor, which can be expressed as a matrix
if $\rho$ is represented as a vector, and $\rho^{t_0}$ is the density matrix at
the initial time $t_0$.
For each eigenstate of the SCPT we define index $l$, which gives the number (including sign) of $4e$ translations
(or $2e$ in the asymmetric case) the state differs from its equivalent state in the central zone.
This defines an operator $\hat L$ and the net current can be expressed as
the time average of $2e\frac{d}{dt}\langle \hat{L}(t)\rangle$. After the Laplace transform this becomes
\begin{gather}
I=2e\lim_{s\rightarrow 0} s^2\langle\hat{L}(s)\rangle.
\label{current}
\end{gather}

Since the transition rates are translationally invariant in $l$ (Appendix), we trace the master equation
with respect to this variable. Labeling the states as $(a,l)$, we include only diagonal entries in $l$ to the master equation  (i.\ e.\ the elements $\rho_{(a,l_1)(b,l_2)}$ are not taken
into account if $l_1\neq l_2$) to be able to do the trace. Since a state inside the central zone can be freely selected
to be any of the equivalent states, the physics lost in this approximation
is usually minimized by minimizing the energy differences of the states
inside a zone. This selection is done separately for each $V$.
The following method can be extended to include also nondiagonal entries
of the density matrix between neighboring zones, etc.,  but according to our numerical simulations
the results are not usually changed if the zones are chosen
as mentioned. An exception is mentioned in section \ref{results}.

Using the translational invariance we can now do the trace over the zones in a Fourier-transform fashion by summing the different zone contributions
as~\cite{zeno}
\begin{gather}
s\rho_{mn}(k,s)-\rho_{mn}^{t_0}(k)\nonumber\\
\equiv\sum_{l_f}[s\rho_{(m,l_f)(n,l_f)}(s)-\rho_{(m,l_f)(n,l_f)}^{t_0}]e^{-ikl_f}\nonumber\\
=\sum_{a,b,l_i,l_f}\tilde\Sigma^{(a,l_i)\rightarrow(m,l_f)}_{(b,l_i)\rightarrow(n,l_f)}(s)\rho_{(a,l_i)(b,l_i)}(s)e^{-ikl_f}e^{ik(l_i-l_i)}\nonumber\\
=\tilde\Sigma^{a\rightarrow m}_{b\rightarrow n}(k,s)\rho_{a,b}(k,s),
\label{main}
\end{gather}
where we have defined the traced transition rate operator
\begin{gather}
\tilde\Sigma^{a\rightarrow m}_{b\rightarrow n}(k,s)=\sum_{l}\tilde\Sigma^{(a,0)\rightarrow(m,l)}_{(b,0)\rightarrow(n,l)}(s)e^{-ikl}.
\end{gather}
By inverting the matrix $s-\tilde\Sigma(k,s)$, the net current can be expressed via the relation
\begin{gather}
\langle \hat{L}(s)\rangle=i{\rm{Tr}}\{\frac{d}{dk}\rho(k,s)\}\mid_{k=0}\nonumber\\
=i\frac{d}{dk}{\rm{Tr}}\{[s-\tilde\Sigma(k,s)]^{-1}\rho^{t_0}(k,s)\}\mid_{k=0},
\end{gather}
and Eq.\ (\ref{current}).

\section{Results}\label{results}

Qualitatively, depending on the parameters of the system,
the $I-V$ characteristics can be divided into three regions
depending on what kind of processes are dominant.
At low voltages $eV\lesssim\Delta$ the
characteristics are $2e$-periodic as there are no strong quasiparticle
tunneling processes. Although
higher-order Cooper-pair tunneling can trigger quasiparticle tunneling by releasing the needed energy $2\Delta$ for creating quasiparticle excitations on both sides of a JJ, this usually occurs at a much lower rate than the decay of the odd quasiparticle state (section \ref{qparticles}).
The $I-V$ characteristics are mainly determined by the coupling of the SCPT to the EE and CF.
At high voltages $eV\gtrsim 4\Delta$ (or $eV\gtrsim 2\Delta$ in the highly asymmetric case) ordinary quasiparticle tunneling
can occur across both the JJs (across the probe) and
the net current increases steeply almost to the normal state value,
which is several orders of magnitude higher than a typical subgap current.
In between is the region where, for example, nonresonant Cooper-pair tunneling can trigger simultaneous quasiparticle tunneling leading to steps in the $I-V$ characteristics.
For applications the most important processes in this region are the
Josephson-quasiparticle (JQP) cycles\cite{fulton,nakamura2,clerk}, where resonant Cooper-pair
tunneling is accompanied by tunneling of quasiparticles.
We focus our analysis mainly on the higher-order resonances in the region $V<\Delta/e$. This is because the characteristics
nearby the usual JQP-cycles are not much different from the results
obtained previously by similar density-matrix approaches\cite{choi1,choi2,clerk,johansson1,johansson2}, as the strong quasiparticle tunneling is
usually behind the broadening of the JQP-cycles overshadowing the effect of the EE and CF ($T_{\rm{CF}}=0$).

\subsection{Asymmetric SCPT interacting with EE}\label{asymSCPT}

We first review the results obtained by the CTM for the asymmetric SCPT~\cite{lt24,oma}.
When represented in the energy eigenbasis of the SCPT, the CTM includes transitions between the diagonal elements of the density matrix
(i.~e.~populations), but neglects the off-diagonal ones.
The suitable form of the SCPT Hamiltonian
in the asymmetric situation, Eq.~(\ref{hamasym}), consists of two parts; the CPB Hamiltonian and the
probe tunneling operator.
The name probe is justified since the current $I(V)$
has peaks (at least) whenever the energy difference
between the ground and an excited state of the CPB is equal to $2eVn$,
the energy released in the tunneling of $n$ Cooper-pairs across the probe.
At such a resonance the probe tunneling
operator mixes the degenerate
states $\vert 0,Q_0'\rangle\vert 0\rangle$ and $\vert j,Q_0'\rangle\vert 2en\rangle$ (the basis is defined in section \ref{model2}) and causes avoided crossings such that the
minimum level separation of the corresponding SCPT eigenstates
is
\begin{gather}
\delta E=E_{J2}\vert\langle j,Q_0'\vert T_{Q}^{\dagger}\vert 0,Q_0'\rangle\vert,
\end{gather}
for the first order resonances ($n=1$).
The state $\vert j,Q_0'\rangle\vert 2en\rangle$ usually decays to the state
$\vert 0,Q_0'\rangle\vert 2en\rangle$ via a photon emission to the EE,
which is then degenerate with $\vert j,Q_0'\rangle\vert 4en\rangle$ and
so on.
Thus the resonance leads to efficient charge transport.
The decay of the CPB states via quasiparticle tunneling across the probe (or the larger JJ) is similar but costs an energy $2\Delta$ so that in resonant situations it occurs
approximately above $V=2\Delta/3e$ ($V=\Delta/e$), see section~\ref{quasieffect}.
The full width at half maximum of the corresponding $I(V)$ peak is determined
by the width of the mixing region which for $n=1$ is approximately $\delta E/e$.
For higher order resonances ($n>1$) also virtual intermediate states
contribute and the splitting [and the $I(V)$ peak width] has to be calculated
separately.
For a more detailed discussion of the CTM in the asymmetric situation see Ref.~\onlinecite{lt24}.

\begin{figure}[tb]
\begin{center}\leavevmode
\includegraphics[width=\linewidth]{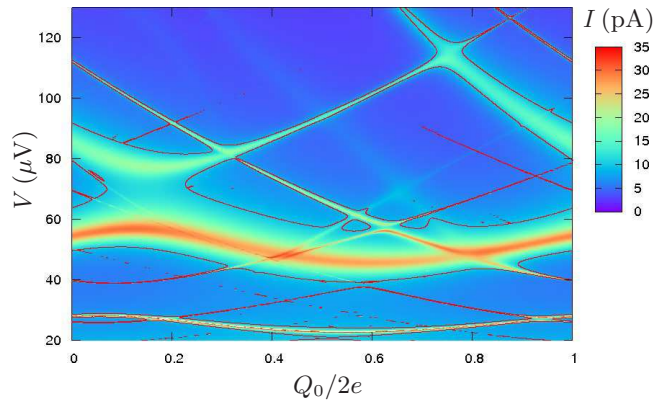}
\caption{
(Color online) A two-dimensional map of the current $I$ as a function of $Q_0$ and $V$
according to the CTM.
A contour line for the current $10$ pA is drawn for clarity.
The resonances originate in the first or higher-order Cooper-pair tunneling
across the probe and simultaneous excitation of the CPB, described by
$H_{\rm{SCPT}}$ (\ref{hamasym}). The resonance positions oscillate as a function of $Q_{0}$ with the period of $2e$ due to the
band structure of the CPB eigenstates. Since $Q_0'=Q_0-C_2V$ (the quasicharge in the CPB Hamiltonian), the
change in $V$ also leads to a change in $Q_0'$ and therefore the band structure is skewed when plotted
as a function of $Q_0$.
The parameters are
$E_{J1}=5E_C=12.5E_{J2}=\Delta/2.2=100$ $\mu$eV, $C_2/C_1=0.21$, $T=50$~mK and $R=50$~$\Omega$.
}
\label{num1}\end{center}\end{figure}

Fig.~\ref{num1} shows the $I(V)$ resonances for
an asymmetric SCPT with $E_{J1}/E_C=5$ calculated numerically using the CTM.
The band structure (quasicharge dependence) of the
eigenstates is evident even though $E_{J1}/E_C>1$.
The strong resonance oscillating in between the voltages $40$~$\mu V$ and $60$~$\mu V$
occurs since the energy released in a tunneling of a single Cooper pair
across the probe matches the CPB excitation energy to the second (first excited) band.
A similar resonance between the voltages $70$~$\mu V$ and $120$~$\mu V$
occurs due to the CPB excitation to the third band.
Their $Q_0$ dependence is in the opposite ``phase'' because
of the band structure. A second-order transition to the third band is located at $40$~$\mu V$ for $Q_0=0$. In the region $V\in 50-70$~$\mu V$ and $Q_0/2e\in 0.4-0.8$ also first-order transitions
from the second to higher bands are distinguishable since the second band
is populated due to the simultaneous $1\rightarrow 2$ resonance.
A second-order transition to the second band is at $28$~$\mu V$ for $Q_0=0$ and a third-order transition to the third band
at $26$~$\mu V$ for $Q_0=0$. The quasiparticle tunneling at these
voltages is very small and can be neglected.

\begin{figure}[tb]
\begin{center}\leavevmode
\includegraphics[width=\linewidth]{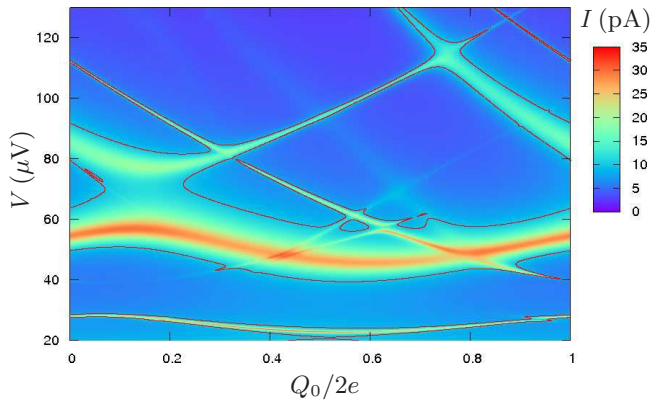}
\caption{
(Color online) A two-dimensional map of the current calculated using the DMA and the same parameters as in Fig.~\ref{num1}.
A contour line for the current $10$ pA is drawn.
As compared to the results obtained by the CTM (Fig.~\ref{num1}), the characteristics are similar but differ for the weak higher-order resonances. Especially, the higher-order resonances between
the first and the third band are barely visible. The first-order resonances are approximately of equal strength
in both models.}
\label{num11}\end{center}\end{figure}

Fig.~\ref{num11} shows the $I(V)$ resonances for
the same system but now calculated using the DMA.
The characteristics are similar, but differ mostly for the weak
higher-order resonances. Especially
the resonances between the first and the third band
have vanished or are significantly weakened in the range from $20$~$\mu V$ to $50$~$\mu V$. The second-order resonance to the first band still persists but its magnitude is reduced.
The peaks due to the first-order resonances are of equal height in both models.
According to our analysis the wash-out of the higher order resonances occurs due to two decohering mechanisms that can be treated independently.
The first one is the level broadening
due to a finite lifetime of the CPB excited bands caused by quantum noise of the EE. The second one is
thermal fluctuations of the voltage across the SCPT (thermal noise).

\begin{figure}[tb]
\begin{center}\leavevmode
\includegraphics[width=\linewidth]{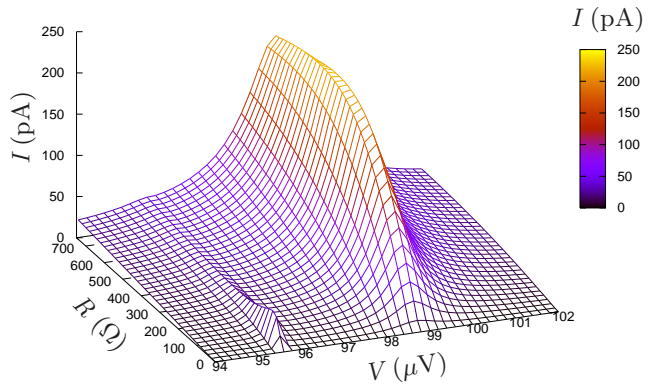}
\caption{
(Color online) The effect of the quantum noise according to the DMA.
A first-order
resonance between the ground and the first excited state (which is a level since the band width is very small) is located at $V\sim 99$~$\mu$V
and a second-order resonance  between
the ground and the second excited state at $V\sim 94.5$~$\mu$V.
With increasing $R$, the resonant transport first
increases $\propto R$ (coherent regime) and finally starts to decrease $\propto 1/R$ (incoherent regime).
The width of the resonance in the incoherent regime is determined by the lifetime of the CPB excited state~\cite{oma} and in the coherent regime by the
splitting of the SCPT eigenstates.
The second-order transition has much smaller splitting and therefore it is more easily washed out by the quantum noise of the EE.
The parameters are $E_{J1}=48.4E_C=500$~$\mu$eV,
$E_{J2}=3$~$\mu$eV, $C_2/C_1=0.1$ and $T=0$. No quasiparticles are included
as their effect is small for $\Delta>200$ $\mu$eV.
}
\label{num4}\end{center}\end{figure}

Fig.~\ref{num4} shows quantitatively the effect of the quantum noise
by increasing the resistance $R$, and at the same time the transition rates since they are $\propto R$ [see Eq.~(\ref{laplace})],
at zero temperature when $E_{J1}/E_C\approx 50$~(Ref.~[\onlinecite{oma}]).
As $R$ is increased,
the transport is first enhanced linearly in agreement with the CTM results
(not plotted). In this region the relaxation is only a small perturbation
as compared to the SCPT eigenstate splittings and the
reduced density matrix is almost diagonal in this basis.
By further increasing $R$, the widening (due to the finite lifetimes) of the CPB excited states starts to
overcome these splittings and the transport saturates and finally slows down, approaching the incoherent limit
$I\propto 1/R$.
This is because during the time evolution the fast relaxation damps the system
to the CPB ground state (which is a superposition of the SCPT eigenstates),
and the density matrix becomes almost diagonal in the CPB basis.
This is a manifestation of the Zeno or watchdog effect~\cite{zeno2,zeno}.
The CPB can still evolve to the excited state but now with a much
lower (incoherent) rate $\propto 1/R$. Therefore the Cooper-pair tunneling
slows down and the net current decreases.
A similar effect occurs when the splitting of the states behind the resonance is lowered by decreasing $E_{J2}$ at a constant $R$.

\begin{figure}[tb]
\begin{center}\leavevmode
\includegraphics[width=\linewidth]{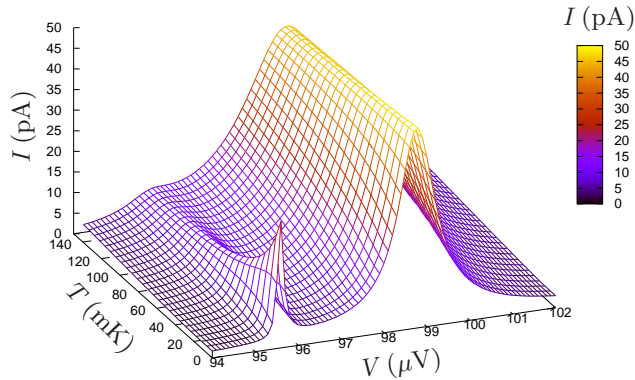}
\caption{
(Color online) The effect of thermal noise on the first and second-order $I(V)$ resonances of Fig.~\ref{num4} with a fixed $R=50$~$\Omega$.
The evolution in the first-order resonance is mainly coherent (linear regime in Fig.~\ref{num4}) and the effect of thermal noise is
fairly small (the area of the peak increases). Qualitatively this is because the bottleneck of the current is the slow relaxation of
the populated excited state. The result is similar as obtained using the CTM (not plotted).
However, the tunneling in the second-order resonance is partially incoherent
and the increase of temperature
tends to broaden (with a constant area) and finally flush the resonance. Qualitatively this is because the bottleneck of the current is the
excitation of the CPB which is strongly perturbed by thermal fluctuations of $V$.
In the CTM the second-order resonance widens but does not substantially lose its height (the area increases) as the temperature is increased.}
\label{num3}\end{center}\end{figure}

Fig.~\ref{num3} shows the effect of thermal noise
on the same system but now with a fixed $R$. It demonstrates that
strong thermal fluctuations also lead to analogous wash-out of certain resonances.
In Ref.\ [\onlinecite{oma}] it was derived that the (first-order)
incoherent Cooper-pair tunneling probability function broadens by
$\Gamma_{2e}=4\pi k_bTR/R_Q$ due to thermal noise (see also section \ref{treshold}).
The DMA produces the same result in the limit of incoherent tunneling and similar effect occurs also for the higher-order resonances.
However, if the Cooper-pair tunneling in the resonance is coherent, the
$I(V)$ peak broadens but does not significantly weaken.
One sees from Eq.~(\ref{interaction}) that the CPB variable $Q$ is protected from the EE's noise by the small $C_2/C_{\Sigma}$ ratio.
This reduction factor does not appear in the $Q_\Sigma$ variable, which causes dephasing (energy level fluctuations) of the tunneled charge states~\cite{shnirman2}, and delivers
thermal noise to the system causing broadening.
In order to see the higher-order effects experimentally, besides minimising the environmental temperature, the (low frequency) resistance of
the voltage line should be minimised also, for example, by nearby electrical components.
Numerical results show that the factor $\Gamma_{2e}$ should not be much larger than the splitting of the relevant SCPT
eigenstate behind the resonance if the Cooper-pair tunneling is incoherent
(broadening due to the quantum noise larger than the splitting).

\subsection{Asymmetric SCPT interacting with CF}

The effect of the quantum $f$-noise caused by the CF is directly linked to the results of the preceding subsection.
The characteristics for this situation can be obtained by simply changing $Q_{\rm{int}}^{\rm{EE}}$, Eq.~(\ref{interaction}), to
$Q_{\rm{int}}^{\rm{CF}}=Q$ (section \ref{CF}) and readjusting the environmental resistivity to a proper value.
In Ref.\ [\onlinecite{nakamura1}] the value $R_{\rm{CF}}=3$~$\Omega$ was found to match the experimental
results and we use this in the simulations. Since the
current is second order in the charge operators, and neglecting the effect of $Q_{\Sigma}$, a similar effect is caused by an EE
with $R=(C_{\Sigma}/C_2)^2R_{\rm{CF}}$. This corresponds to $R\approx 100$~$\Omega$ if $C_2/C_{\Sigma}\approx 1/6$ (Fig.~\ref{num1}), and  $R=300$~$\Omega$ if $C_2/C_{\Sigma}\approx 1/10$ (Fig.~\ref{num4}). Therefore the effect of the CF is usually dominant
for a small $C_2/C_{\Sigma}$ ratio.

\begin{figure}[tb]
\begin{center}\leavevmode
\includegraphics[width=\linewidth]{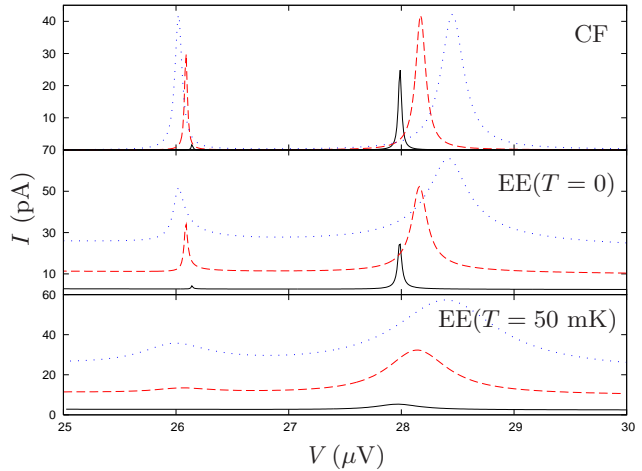}
\caption{
(Color online) The current nearby two
higher-order resonances of Fig.~\ref{num11} for $Q_0=0$ while changing $E_{J2}$ for three different environments.
The solid lines correspond to $E_{J2}=5$ $\mu eV$,
the dashed lines to $E_{J2}=10$ $\mu eV$ and the dotted lines to $E_{J2}=15$ $\mu eV$.
The calculation for the CF (and no EE) is made using an effective resistance $R_{\rm{CF}}=(C_2/C_{\Sigma})^250$~$\Omega$,
the calculation for the EE (and no CF) with $R=50$ $\Omega$, $T=50$ mK but also with $T=0$, i.~e.~when no thermal noise is present.
The zero temperature current for the EE and CF are similar,
except that the latter has no nonresonant ``background'' current $\propto E_{J2}^2R/V$. The thermal noise of EE with $T=50$ mK broadens the resonances and flushes them in the case of small $E_{J2}$.
}
\label{cf}\end{center}\end{figure}

Numerical calculations show that near resonances the $I-V$ characteristics
are approximately the same as in the case of an EE with $R=R_{\rm{CF}}(C_{\Sigma}/C_2)^2$ and $T=0$,
see Fig.~\ref{cf}. This is a consequence of the fact that
the excited states of the CPB decay only via the operator $Q$ which also gives the dominant contributions to the
matrix elements between the SCPT eigenstates in resonant situations.
However, the nonresonant current $\propto E_{J2}^2R/V$~\cite{lt24,ingold} is not present in the case of CF since it is related to the fluctuations of $Q_{\Sigma}$.
This is the main difference between the $I-V$ curves and therefore the low bias behaviour distinguishes between the
two different sources of noise.

If the CF and EE are present simultaneously, the thermal noise of the EE can greatly contribute to the rates due to the CF.
Contrary to the case of quasiparticle tunneling (section \ref{treshold}), this effect is already included by summing up two different transition rate contributions (from the EE and CF),
following from the fact that the Cooper pair tunneling is treated nonperturbatively.
The thermal noise leads to similar broadening of the resonances
as in the case of the EE, but is now characterized by
the resistivity of the EE through the corresponding broadening
factor. Therefore, even if the EE would not flush the
weak resonances due to a small $R$ or $C_2/C_{\Sigma}$ (slow relaxation due to the EE),
the CF can drive the tunneling across the probe incoherent, which then accompanied by moderate thermal fluctuations of EE can lead to wash-out of the higher-order
resonances. Similarly, thermal fluctuations of the CF and quantum noise
due to the EE can lead to the same effect.

\subsection{Asymmetric SCPT with quasiparticle tunneling}\label{quasieffect}

The quasiparticle tunneling becomes significant in resonant situations when $V\gtrsim 2\Delta/3e$ for $E_{J1}/E_C\gg 1$. This is since
the decay of a resonant CPB state (releasing energy $\sim E_{\alpha}-E_0=2eV$) via quasiparticle tunneling across the probe ($\sim eV$) releases the total energy $\sim 3eV>2\Delta$. The tunneling across the larger JJ
is more intense but releases no energy, and occurs therefore approximately
at voltages
$V\gtrsim\Delta/e$. These relations are only
qualitative as the true threshold voltages depend on the details of the
energy level structure (band structure if $E_{J1}/E_C\sim 1$) and usually are between $\Delta/2e<V<2\Delta/3e$.
In analogy to the EE or CF,
if the decay rate due to quasiparticle tunneling becomes larger than the splitting of the SCPT eigenstates
behind the resonance, the Cooper-pair
tunneling turns incoherent and charge transport slows down (Fig.\ \ref{num4}). For voltages well below $\Delta/e$ a more destructive
effect usually is the simultaneous $e$-switching of the quasicharge $Q_0'$, which then leads to an off-resonance situation. The system
returns to the resonance, for example by a decay of the odd
parity state of the island, but at a much lower rate than typically at
resonance.
However, the switching
has no effect in the case of a double-resonance point, where the resonance
occurs simultaneously in both parity states (for example when $Q'=e/2$), or when
the relevant states of the CPB have practically no quasicharge dependence ($E_{J1}/E_C\gg 1$).

\begin{figure}[tb]
\begin{center}\leavevmode
\includegraphics[width=\linewidth]{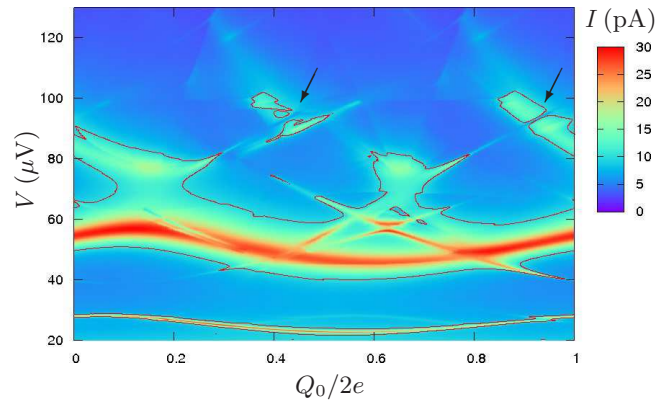}
\caption{
(Color online) A two-dimensional map of the current calculated
using the DMA and the same parameters as in Fig.~\ref{num1}, except $\Delta=150$ $\mu$eV.
A contour line for the current $10$ pA is drawn. The resonances below $V\sim 80$ $\mu$V
maintain $2e$-periodicity as the quasiparticle tunneling is very small at these
voltages. The resonances above this start to show $e$-periodicity as
the decay process releases enough energy for a Cooper pair to break and the quasiparticle
to tunnel across the probe. The arrows point to the double-resonance points of the third band (see text). All quasiparticle tunneling thresholds have been broadened by the factor $\Gamma_{2e}$ (section \ref{treshold}).
}
\label{qkuva}\end{center}\end{figure}

Fig.~\ref{qkuva} shows the $I-V$ characteristics of the
same system as considered in Figs.\ \ref{num1} and \ref{num11} but now reducing the
superconducting gap to
$\Delta=150$ $\mu$eV and using the DMA. At low voltages the characteristics
are $2e$-periodic in $Q_0$ (as before) but approximately
above $\Delta/2e$, start to show $e$-periodicity as the quasiparticle
tunneling increases.
The $I-V$ characteristics due to the (first-order) excitation to the third band are now modified in comparison with
Fig.~\ref{num11}. The peak due to the excitation to the bottom of the third band is still present but otherwise the transition is visible only near the double resonance points (pointed by the arrows). The characteristics in this region are unchanged when using the CTM (not plotted) which indicates that
the Cooper-pair tunneling is mainly coherent. For $\Delta=100$ $\mu$eV the tunneling in this region becomes partly incoherent
as the quasiparticle tunneling across the larger JJ also starts to relax the CPB.
In this case only the double resonance point is highlighted.

\subsection{Symmetric SCPT}\label{symSCPT}

\begin{figure}[tb]
\begin{center}\leavevmode
\includegraphics[width=\linewidth]{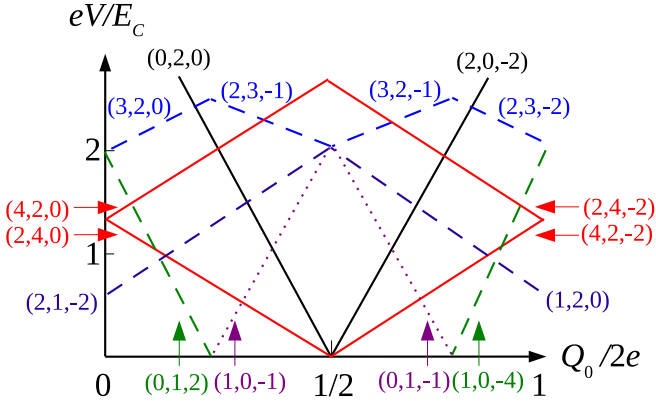}
\caption{
(Color online) Some Cooper-pair tunneling resonances (solid lines), quasiparticle-tunneling thresholds (dashed lines) and the odd-parity-state decay thresholds (dotted lines) in a symmetric SCPT with $E_J<E_C$.
The quasiparticle thresholds are calculated assuming $\Delta=2E_c$.
}
\label{firstreso}\end{center}\end{figure}

Let us now consider resonant tunneling in the symmetric SCPT. We focus on the case
$E_{J}<E_C$ since our choice of the basis (section \ref{model1}) is not optimal for numerical simulations
in the opposite situation. Also the results of the preceding subsections
are less valid for $E_{J}\gg E_C$, where
the nonresonant background current (which is $\propto E_J^4$ in the
symmetric case)
takes a more dominant role in the $I-V$ characteristics.
Since $E_J<E_C$,
the regions of resonant single-Cooper-pair processes can be found analytically
by demanding the degeneracy of the charging energies [the first two terms on the r.~h.~s.~of Eq.~(\ref{hamsym})] before and after
elementary tunneling processes~\cite{nakamura2}. The
process where the charge $le$ tunnels across the left JJ and $re$ across the right JJ (both to the positive direction) is resonant when
\begin{gather}
V=\frac{4E_C}{e(l+r)}\left[\frac{(l-r)^2}{2}+\left(\frac{Q_0}{e}+n\right)(l-r)\right]\label{mainsset},
\end{gather}
where $en$ is the initial island charge. This releases energy for larger values of the voltage and if
it includes quasiparticle tunneling (i.~e.~$l$ or $r$ is an odd number and the process is not the decay of the odd parity state)
the term $4\Delta/e(l+r)$  has to be added to
the r.~h.~s.~of Eq.~(\ref{mainsset}).
After such a quasiparticle process the charge on the island changes parity, which effectively means shifting $Q_0$ by $e$. Efficient charge transport including quasiparticles requires cycles of rapid processes in both parity states.
In the following we label the tunneling processes as $(l,r,n)$.
The locations of some resonances and thresholds are drawn in Fig.~\ref{firstreso}.

\begin{figure}[tb]
\begin{center}\leavevmode
\includegraphics[width=\linewidth]{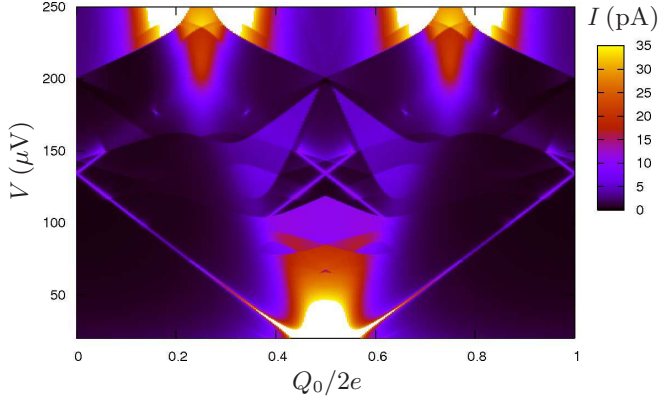}
\caption{
(Color online) A two-dimensional map of the current across the symmetric SCPT according to the DMA.
Some of the resonances and thresholds are identified in Fig.~\ref{firstreso}.
In the white regions the current exceeds the plotting range.
All quasiparticle tunneling thresholds have been broadened by the factor $\Gamma_{2e}$.
The parameters are $E_C=3.3E_J=\Delta/2=100$~$\mu$eV,
$R=50$~$\Omega$, $T=50$~mK and $R_{\rm{CF}}=3$~$\Omega$.
}
\label{sym1}\end{center}\end{figure}
Fig.~\ref{sym1} shows the $I-V$ characteristics for $E_J/E_C=0.3$ and typical EE and CF calculated using the DMA.
At low voltages and $Q_0\sim e$ the current is enhanced
as the processes $(0,2,0)$ and $(2,0,-2)$ (see also Fig.~\ref{firstreso}) with matching initial and final states
produce an efficient charge transport cycle. When quasiparticle tunneling is not included, the area of the enhanced current is V-shaped, fenced by the resonance lines, and extends above the plotted region with slowly decreasing current.
In this case the shape of the $I-V$ curve (for a fixed $Q_0$) is more a step than a Lorentzian.
This reflects that, in contrast to the asymmetric case, the single-Cooper-pair tunneling and simultaneous photon emission to the EE or CF
is a strong process above its threshold.
Including the quasiparticle tunneling changes the characteristics from the smooth decrease to steps where quasiparticle tunneling, in most cases accompanied by
tunneling of one or more Cooper pairs, becomes possible. In the step the current 
can also be reduced if the new distribution does not support the charge transport processes below the threshold.

Strong higher-order resonances at low voltages are the processes $(2,4,0)$ and $(4,2,-2)$.
They are visible in Fig.~\ref{sym1} as straight lines and are not washed out due to thermal fluctuations of the EE since their minimum splitting~\cite{brink1} $\approx 81E_J^3/1280E_C^2\approx 0.7$ $\mu$eV is not dominated
by the thermal broadening factor $\Gamma_{2e}\approx 0.4$ $\mu$eV. Also the quantum noise and the quasiparticle tunneling are not too
intense, except for $eV>3E_C/2$ when
ordinary quasiparticle tunneling becomes possible for one of the resonant states. In Fig.~\ref{firstreso} this corresponds to, for example, the crossing point of the processes
$(4,2,0)$ and $(0,1,2)$.
Above this the CTM current increases
but the DMA current weakens significantly due to a strong Zeno effect.
The resonance is also seen in the experiments~\cite{joyez,billangeon}, but disappearing even before the onset of quasiparticle tunneling, perhaps indicating a more pronounced subgap quasiparticle current.
The double-resonance points $eV=2E_C$ and $Q_0/2e=0.25$ or $0.75$ are the locations of the double JQP-cycles but for this
choice of parameters the ordinary quasiparticle tunneling is not possible (the points are below $2\Delta-E_C$) and the transport occurs via higher-order processes. 
The thresholds above these resonances are due to quasiparticle tunneling accompanied by tunneling of two Cooper-pairs. Several low-order processes are behind the strongly enhanced current just above the double-resonance points.

Above the range of Fig.~\ref{sym1} at $Q_0=0$ and $eV=4E_c$, the first-order processes $(2,0,0)$ and $(0,2,0)$ are resonant simultaneously but with nonmatching states. The CTM produces a strong current
peak through higher-order processes~\cite{brink1,brink2} but in the DMA the current is only slightly
enhanced as the strong quasiparticle current destroys the coherence
of the weak higher-order process. Since many resonances occur simultaneously
at this point the DMA needs to be done by considering also the ``neighbouring''
nondiagonal states in the zone parameter (section \ref{tracing}) and by a careful choice of the states that belong to the same zone. The current nearby the ordinary
JQP-cycles is also reduced due to intense quasiparticle tunneling.
Actually, if strictly using the Ambegaokar-Baratoff~\cite{tinkham,ambegaokar} values for the Josephson coupling
energies (with a typical energy gap for an aluminum film), this is always the case for a symmetric SCPT
since the quasiparticle resistance and the Josephson coupling energy cannot be varied independently (for a constant energy gap). This limitation does not apply to an asymmetric SCPT~\cite{nakamura2}.

Finally, we have studied the effect of the CF alone and compared to the case of the EE.
Since $Q_{\rm{int}}^{\rm{EE}}=\bar Q/2$,
it seems that the EE does not induce quasicharge fluctuations ($\propto Q$), which mainly cause the relaxation in the asymmetric case. Thus qualitative
differences could be expected in the case of the CF. This is, however, misleading since this kind of immunity is true only for fully symmetric resonances.
In the asymmetric basis, which
can be used in most resonant situations for the evaluation of the current as well, they exist with a capacitive shielding
$(C_2/C_{\Sigma})^2=1/4$.
Numerical results show again that the CF lead to similar $I-V$ characteristics as the EE with $R=(C_{\Sigma}/C_2)^2R_{\rm{CF}}=4R_{\rm{CF}}$ and $T=0$.  This relation
tells however that an EE with $R=50$~$\Omega$ is
qualitatively about four times more destructive than $R_{\rm{CF}}=3$~$\Omega$, i.~e.~the noise of the EE usually dominates the noise of the CF in the
case of symmetric SCPT.
Special situations are the regions where similar tunneling processes
across both of the junctions are resonant simultaneously, where the effect of the CF vanishes.

\section{Conclusion}\label{conclusion}

We have analyzed typical decoherence mechanisms present in mesoscopic superconducting devices through their effect on the $I-V$ characteristics of a voltage-biased SCPT. We have shown that each of the environments leave different
traces to the characteristics on which ground they can be identified. We have also shown that the higher-order resonances
obtained by the CTM tend to be washed out by
quantum or thermal noise supplied by the environment.
This explains why only few of them have been detected in the
experiments.
In order to see other resonances, the
relevant noise sources causing the wash-out should be filtered as well as possible.
Theoretically one can calculate the minimum splitting
of the SCPT eigenstates causing the resonance and
obtain analytic relations for preventing the wash-out.
In simple terms the splitting should not be
essentially smaller than the decay rate due to the EE, CF or the quasiparticle tunneling, or the resonance is not seen due to the Zeno-effect. If it is smaller, but not essentially smaller,
the thermal noise of the EE, described by a similar broadening factor, tends to flush the resonance.
The effect of the CF's thermal noise, which is usually of $1/f$-type, was not studied
but could also lead to a similar effect.
Finally, if quasiparticle tunneling exists (but is not too intense), the resonance lines are usually only seen near double-resonance points,
because in other regions the quasiparticle tunneling switches the system to an off-resonance situation slowing down the charge transport.

\acknowledgments
We thank M.~Devoret, S.~Girvin, P.~Hakonen and R.~Lindell for useful discussions.
This work was financially supported by the National Graduate School in Materials Physics, by the Academy of Finland, and by the Finnish Academy of Science
and Letters (Vilho, Yrj\"o and Kalle V\"ais\"al\"a Foundation).

\appendix

\section{Calculation of the transition rates}

Here we give rules for calculating the generalized transition rates
between the SCPT's density matrix entries due to the coupling with the EE or CF. The derivation of the
diagram rules is based on a series expansion of the system's time evolution operator in the interaction picture~\cite{keldysh1,keldysh2,keldysh3}.
We also analyze the renormalization, translation invariance
and validity of the higher-order calculation.

\subsection{Interaction with the EE and CF}

The interaction between the SCPT and the EE/CF is described by ``environmental points'' located at arbitrary branches
and times in the Keldysh diagram and are pairwise linked to each other by  ``environmental lines''. The effect of the
renormalization term is described by ``RN pairs''. A RN pair is located at any branch and time but the points forming the pair have the same branch and time. The contributions from the EE/CF and from the renormalization
have to be calculated to the same order in the sense that one RN pair
is equivalent to one environmental line. The generalized transition rate
$\tilde\Sigma^{a\rightarrow m}_{b\rightarrow n}(t-t')$ is a sum of all
irreducible diagrams starting from $t'$ and ending to $t$,
meaning that a vertical line between times $t'$ and $t$ always cuts an environmental line. The rules for constructing the transition rate corresponding to a diagram
are the following.
\begin{enumerate}
\item Each point or RN pair in the upper branch produces a factor $-i/\hbar$ and in the
lower branch a factor $i/\hbar$.
\item Each EE/CF line contributes a factor $\alpha(t_2-t_1)^{(*)}$, where
the complex conjugation takes place if the left end of the line, corresponding to the smaller time $t_1$, is located at the lower branch, and each RN pair a factor $1/2C_{\rm{int}}$.
\item Each point produces a factor
$\langle f\vert Q_{\rm{int}}\vert i\rangle e^{i(\tilde t-t')(\omega_f-\omega_i)}$ where $\vert i\rangle$ is the entering state, $\vert f\rangle$ the leaving state with respect to the direction of the branch, $\hbar\omega_j$ an
eigenenergy of the SCPT's eigenstate $\vert j\rangle$ and $\tilde t$ the timing of the point.
\item An integration over the timings
of EE/CF points or RN pairs, located between the times $t'$ and $t$,
is made and the result is multiplied by $e^{i(t-t')(\omega_n-\omega_m)}$.
\end{enumerate}

\begin{figure}[tb]
\begin{center}\leavevmode
\includegraphics[width=\linewidth]{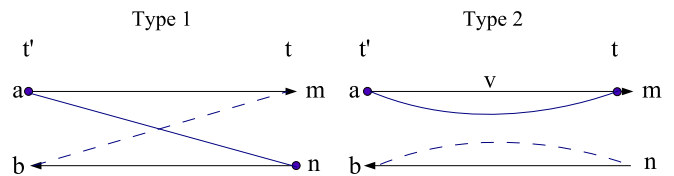}
\caption{
(Color online) The second-order irreducible diagrams describing interaction with the environment. The dashed lines represent the mirror diagrams. Their contribution is the complex conjugate
of the original ones with switched $a\leftrightarrow b$ and 
$m\leftrightarrow n$ (a mirror rule). The arrows point to the directions of time in the two branches.
}
\label{graafi}\end{center}\end{figure}

The second-order irreducible diagrams describing coupling with the EE/CF
are shown in Fig.~\ref{graafi}.
From these one obtains the transition rates (\ref{1order}). Only one RN pair can exist in a second-order graph,
with no accompanying EE/CF points. Thus the only irreducible RN diagram is the one which starts and ends at the same time $t$
producing the lowest order renormalization operator $i/\hbar[\cdot,Q_{\rm{int}}^2/2C_{\rm{int}}]$. Note that according to the rules the exponential
factor of the first-type solid line diagram in Fig.~\ref{graafi}, for example, is
$e^{i(t-t')(\omega_b-\omega_m)}$. The Markov
approximation of the master equation (used in section \ref{lindblad}) would effectively add an extra factor 
$e^{i(t-t')(\omega_a-\omega_b)}$ to the transition rates, as one drops out
the history of the density matrix from the equations. Therefore the
exponential factor of this diagram in the BM
approximation is $e^{i(t-t')(\omega_a-\omega_m)}$ i.~e.~the energy release in the upper branch process (and the lower branch in the case of the mirror diagram).
We have tested that our numerical results are unaffected by the Markov approximation in the cases of EE and CF.

\subsection{Laplace transform and renormalization of the transition rates}

The Laplace transform (or time averaging) of the second-order transition rates
can be reduced to an integral of the form
$\int_{0}^{\infty}dte^{i(\omega+is)t}\alpha(t)$, or in the fourth-order
$\int_{0}^{\infty}dt\int_{0}^{t}dt_2\int_{0}^{t_2}dt_1e^{i(\omega+is)t+i\omega_1t_1+i\omega_2t_2}\alpha(t-t')\alpha(t_2-t_1)$, from which
all the elements of the transformed transition-rate tensor can be constructed.
At least up to the fourth order, the integrals can be evaluated
analytically
by first performing the time integration, then taking the limit $s\rightarrow 0$ and, if needed, using the residue theorem. The second-order integration gives
\begin{gather}
\lim_{s\rightarrow 0}\int_{0}^{\infty}e^{i(\omega+is)t}\alpha(t)dt=\frac{\hbar\omega}{1-e^{-\beta\hbar\omega}}{\rm{Re}}[Z_t(\omega)]\nonumber\\
-i\frac{\hbar\omega_c}{2}{\rm{Re}}[Z_t(\omega)]+i\frac{\hbar\omega}{2\pi}{\rm{Re}}[Z_t(\omega)]\tilde\Psi(\omega),
\label{laplace}
\end{gather}
where
\begin{gather}
\tilde\Psi(\omega)=\Psi\left(1-\frac{\hbar\beta\omega_c}{2\pi}\right)+\Psi\left(1+\frac{\hbar\beta\omega_c}{2\pi}\right)\nonumber\\
-2{\rm{Re}}[\Psi\left(1+i\frac{\hbar\beta\omega}{2\pi}\right)],
\label{psi}
\end{gather}
and $\Psi(x)$ is the digamma function. For this expression we have assumed that $\hbar\beta\omega_c/2\pi=n+1/2$, where $n$ is a large integer.
The first term on the r.~h.~s.\ of
Eq.~(\ref{laplace}) is the usual fluctuation spectrum of the EE or CF and describes
decoherence via emission or absorption of photons. The last two terms are purely imaginary. To a good approximation they produce a constant
$-i\hbar/2C_{\rm{int}}$ at usual frequencies which are well below $\omega_c$.
They induce coherent oscillations and reflect the effective potential caused by the EE or CF. In comparison, the
renormalization $L_N$ in Eq.~(\ref{density1}) contributes through the same, but opposite factor
$i\hbar/2C_{\rm{int}}$. As this enters $\tilde\Sigma$ through the same matrix elements as obtained from the second-type diagrams in Fig.~\ref{graafi}, there is cancellation of the imaginary terms.
The constant complex part also vanishes in the sum of the first-type diagrams,
because the mirror diagrams contribute by complex conjugated terms.

\subsection{Validity of the expansion in the case of the EE}

By looking at Eq.~(\ref{dis}) one can see a problem:
as charge starts
flowing across the SCPT, $Q_{\rm{int}}^{\rm{EE}}$ starts to increase, and as $q$ does not, the charging energy in Eq.\ (\ref{dis})
starts to increase also. So qualitatively speaking, in the exact product-state
calculation the current should stop after some
time the decoupling is made, in order to conserve the energy. Also
the transition rates due to (higher-order) processes that include large changes in the feed charge should be suspicious since the bath does not ``follow'' the state of the system.
This is clearly not what happens
in a real physical situation. The effects following from
the assumption of a static EE restricts the usage of the real-time diagrammatic technique for this problem.

The extra imaginary term in Eq.~(\ref{laplace}), dependent on $\tilde\Psi(\omega)$, is not
a constant and induces (small) spurious dynamics. This indicates that
the effective potential felt by the subsystem in this treatment is not exactly $Q_{\rm{int}}^2/2C_{\rm{int}}$.
However, in the formal solution of Heisenberg equations of motion the embedding of the bath (change in the effective potential) vanishes exactly~\cite{weiss}.
In the second-order calculation the term $\tilde\Psi(\omega)$ does not contribute in the diagonal terms,
as the complex part vanishes, and the contribution through nondiagonal terms is small.
Still diagrams that posses terms like
$\langle i\vert Q_{\rm{int}}^{\rm{EE}}\vert i\rangle$, which depend on the ``position'' in the W-S ladder are a source of small
translational variance, but their effect seems to be small as long as
$\vert\langle Q_{\rm{int}}^{\rm{EE}}\rangle\vert/e\approx 1$.
The effect can be studied by shifting the central zone.

In the fourth-order calculation the previous change in the effective potential
and the translational variance seem to enhance each other.
This occurs because after each integration the corresponding RN diagrams do not remove the embedding of the
EE completely but leave behind the discussed extra term. This is then
enhanced by terms like $\langle i\vert Q_{\rm{int}}^{\rm{EE}}\vert i\rangle$,
which can exist two times in diagrams that
describe decay, producing strong translational variance and dissipation, which is essentially larger than in the first-order calculation.
Also nondiagonal elements suffer from the same effect.
If one now removes the corresponding spurious terms after each integration from the calculation ``by hand'', the discussed behaviour vanishes and the final correction is
small compared to the first order calculation, as expected for $R/R_Q\ll 1$.
So for analysing the effect of the EE in the case of larger $R$, where the
higher order effects should become dominating, the model cannot be used
in this form.

\end{document}